\documentclass{nature}

\usepackage{amsmath,amssymb}
\usepackage{xcolor}

\usepackage{caption}

\usepackage{bm}
\usepackage{braket}
\usepackage{amssymb}
\usepackage{textcomp}

\usepackage{bbold}
\usepackage{mathdots}
\usepackage[thicklines]{cancel}
\usepackage{lineno}

\usepackage{hyperref}
\hypersetup{%
	breaklinks=true,%
	citecolor=blue,%
	colorlinks=true,%
	linkcolor=[RGB]{0,0,200},%
	urlcolor=blue
}	

\usepackage{makecell}

\title{Observation of an isolated flat band in the van der Waals crystal NbOCl$\bm{_2}$}

\author{Changhua Bao$^{1}$, Vincent Eggers$^{1}$, Manuel Meierhofer$^{1}$, Jakob Helml$^{1}$, Lasse M{\"u}nster$^{1}$, Suguru Ito$^{2}$, Leon Machtl$^{1}$,  Sarah Zajusch$^{2}$, Giacomo Inzani$^{1}$, Ludwig Wittmann$^{1}$, Marlene Liebich$^{1}$, Robert Wallauer$^{2}$, Ulrich H{\"o}fer$^{1,2,*}$ and Rupert Huber$^{1,*}$}

\usepackage{graphicx}
\makeatletter
\let\saved@includegraphics\includegraphics
\AtBeginDocument{\let\includegraphics\saved@includegraphics}

\makeatother

\begin{document}

\maketitle
\begin{affiliations}
\item Department of Physics and Regensburg Center for Ultrafast Nanoscopy (RUN), University of Regensburg, 93040 Regensburg, Germany. 
\item Department of Physics, Philipps-University of Marburg, 35032 Marburg, Germany. \\
$^*$e-mail: hoefer@physik.uni-marburg.de, rupert.huber@physik.uni-regensburg.de
\end{affiliations}


\newpage

\begin{abstract}
{
\bf Abstract\\
Dispersionless electronic bands lead to an extremely high density of states and suppressed kinetic energy, thereby increasing electronic correlations and instabilities that can shape emergent ordered states, such as excitonic, ferromagnetic, and superconducting phases. A flat band that extends over the entire momentum space and is well isolated from other dispersive bands is, therefore, particularly interesting. Here, the band structure of the van der Waals crystal NbOCl$_2$ is revealed by utilizing photoelectron momentum microscopy. We directly map out an electronic band that is flat throughout the entire Brillouin zone and features a width of only $\sim$100 meV. This band is well isolated from both the conduction and remote valence bands. Moreover, the quasiparticle band gap shows a high tunability upon the deposition of caesium atoms on the surface. By combining the single-particle band structure with the optical transmission spectrum, the optical gap is identified. The fully isolated flat band in a van der Waals crystal provides a qualitatively new testbed for exploring flat-band physics.}

\end{abstract}

\renewcommand{\thefigure}{\textbf{Fig.~\arabic{figure}}}
\setcounter{figure}{0}

\renewcommand{\thetable}{\textbf{Table~\arabic{table}}}

When the energy of crystal electrons is independent of their momenta, the corresponding band dispersion becomes flat and the kinetic energy is strongly quenched while the density of states diverges\cite{sutherland1986localization,mielke1992exact,wu2007flat,suarez2010flat,leykam2018artificial}. These exotic properties result in instabilities facilitating emergent ordering phenomena as well as new bound and strongly correlated states, such as ferromagnetism\cite{mielke1993ferromagnetism,sharpe2019emergent}, excitons\cite{wilson2021interlayer,dirnberger2023magneto,liebich2025controlling,shao2025magnetically}, and superconductivity\cite{nandkishore2012chiral,cao2018unconventional}.  Flat bands can be realized by a large superlattice-induced gap opening, which divides the electronic dispersion into narrow bands as demonstrated in charge density wave materials\cite{zwick1998spectral} and moir\'e systems\cite{bistritzer2011moire,cao2018unconventional,cao2018correlated}. Flat bands can also be realized by destructive quantum interference in Kagome lattices\cite{mielke1992exact, kang2020topological,han2021evidence,sun2022observation}, Lieb lattices\cite{lieb1989two,slot2017experimental}, and intercalated transition metal dichalcogenides\cite{peng2025flat}. However, the flat bands in these systems are either only partially flat or they are connected with or lie close to other dispersive bands, limiting their potential to host certain correlated phases and to explore flat-band physics. Thus, an ideal flat band that extends over the entire Brillouin zone and is well isolated from other dispersive bands is highly desired.

Niobium oxide dichloride (NbOCl$_2$) is a newly emerging van der Waals material suggested to form such an ideal flat band. Recently, it has attracted great research interest due to many fascinating electronic and optical properties, including room-temperature in-plane ferroelectricity\cite{jia2019niobium,liu2023ferroelectricity,huang2024ferroelectric,ding2024strain} and highly efficient second harmonic generation (SHG)\cite{guo2023ultrathin,abdelwahab2023highly,ye2023manipulation} in both bulk and monolayer crystals, colossal optical anisotropy\cite{guo2024colossal} and a promising application as a quantum light source\cite{guo2023ultrathin,kallioniemi2024van,guo2024polarization} in tens-of-nanometer-thick flakes. The flat band has been predicted to be close to the Fermi level and well isolated from both the conduction band (CB) and the other valence bands (VBs)\cite{jia2019niobium}. Due to the large effective mass associated with the flat band, a record high exciton binding energy of 800~meV has been predicted even in the three-dimensional (3D) bulk crystal\cite{guo2023ultrathin}, which may also play an important role in the observed high efficiency in SHG\cite{ding2024exciton} and excellent performance for quantum light sources\cite{PhysRevLett.132.246902}. Spin splitting and magnetism have been anticipated by tuning the Fermi level across the flat band\cite{mohebpour2024origin}.

Moreover, NbOCl$_2$ is also an ideal system for exploring the unique role of flat bands in lightwave electronics\cite{borsch2023lightwave,reimann2018subcycle} and light-induced band engineering\cite{bao2022light,wang2013observation,zhou2023pseudospin,ito2023build}. For example, the lightwave-driven intraband electronic transport can be suppressed due to the large effective mass, leading to a novel interband-polarization dominated high harmonic generation\cite{li2025interband}. However, the band structure of NbOCl$_2$ has not been experimentally observed yet. Fundamental electronic properties for exploring exciton and correlation physics, including the band gap and carrier doping level, remain elusive (Supplementary Table 1).

Here, we reveal the band structure of the van der Waals crystal NbOCl$_2$ by utilizing photoelectron momentum microscopy. A flat band extending over the entire Brillouin zone is directly resolved with a narrow width of 100 meV and large separations from the CB and remote VBs. The quasiparticle band gap is revealed by $in$-$situ$ surface electron doping, exhibiting high tunability. Moreover, the optical transition corresponding to the optical gap is revealed based on the experimental band structure.
\begin{figure*}[htbp]
	\centering
	\includegraphics[]{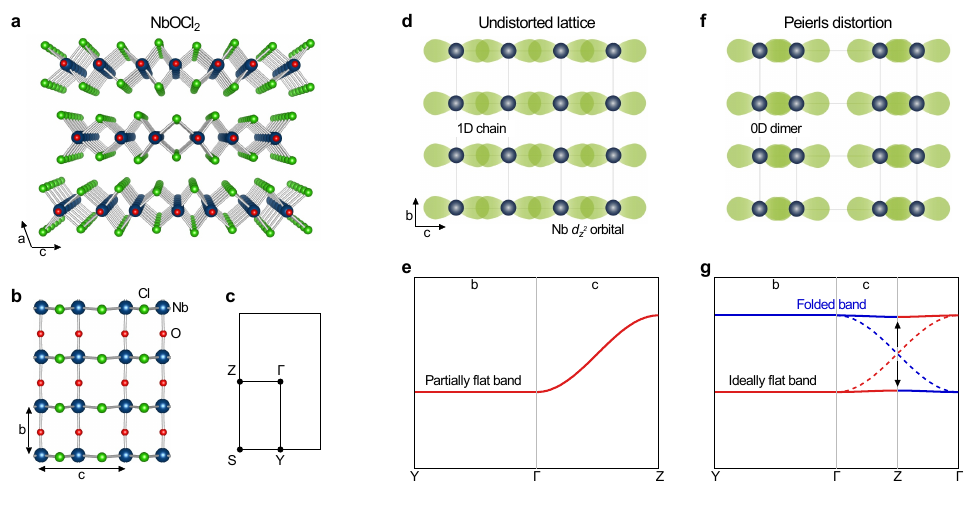}
	\caption{{\bf $\bm{|}$ NbOCl$_2$ as a potential host of flat bands.} \textbf{a, b}, Crystal structure of the van der Waals crystal NbOCl$_2$ from side and top views.  \textbf{c}, The surface Brillouin zone with labeled high-symmetry points. \textbf{d, e}, Ideal square lattice of Nb atoms without Peierls distortion (\textbf{d}). The green shadows represent the Nb 4$d_z^2$ orbitals, whose overlap results in a partially flat band as schematically shown in \textbf{e}. \textbf{f, g}, Realistic rectangular lattice of Nb atoms with a Peierls distortion (\textbf{f}) and corresponding ideal flat band as schematically shown in \textbf{g} owing to the band folding and gap opening. The dashed curves are dispersions without gap opening.
}
\label{F1}
\end{figure*}

\noindent\textbf{Results}\\
\noindent\textbf{Determination of the band structure}\\
NbOCl$_2$ exhibits a layered crystal structure with weak interlayer van der Waals coupling along the a-axis\cite{hillebrecht1997structural,jia2019niobium} as illustrated in \ref{F1}a.  The Nb atoms form an in-plane rectangular lattice featuring a Peierls distortion along the c-axis\cite{jia2019niobium} (\ref{F1}b), whose 4$d$ orbitals dominate the flat band and the CB\cite{ye2023manipulation}. In particular, the flat band is mainly formed by Nb 4$d_{z^2}$ orbitals, which exhibit a lobed shape along the c-axis, as illustrated in \ref{F1}d. The origin of the flatness of the band can be intuitively understood in the following way. Starting from an ideal square lattice without Peierls distortion, the $d_{z^2}$ orbitals overlap with each other and form one-dimensional (1D) chains along the c-axis, which are decoupled along the b-axis (\ref{F1}d). This naturally leads to a dispersive band along the c-axis ($\Gamma$-Z, as labeled in \ref{F1}c), which is dispersionless along the b-axis ($\Gamma$-Y) as schematically illustrated in \ref{F1}e. The partially flat band has been reproduced by first-principles calculations without Peierls distortion\cite{guo2024colossal}. In the realistic distorted lattice (\ref{F1}f), two Nb atoms approach each other to form a zero-dimensional (0D) dimer, which results in the decoupling of the orbitals along both b- and c-axis. In the band structure picture, the doubling of the unit cell folds the band along the c-axis ($\Gamma$-Z) back (blue dashed curves in \ref{F1}g) and opens a gap at the band crossing point due to the inter-dimer interaction, as indicated by the black arrows. This process turns the partially flat band into an ideal flat band. It also leads to a second flat band at higher energy, which is evidenced by the appearance of dispersionless states at an energy of $\sim$4~eV above the first flat band in first-principles calculations\cite{ye2023manipulation}. Therefore, flat bands in NbOCl$_2$ originate from the interplay between the highly anisotropic shape of the Nb 4$d_{z^2}$ orbitals and a Peierls distortion of the lattice.

\begin{figure*}[htbp] 
	\centering
	\includegraphics[]{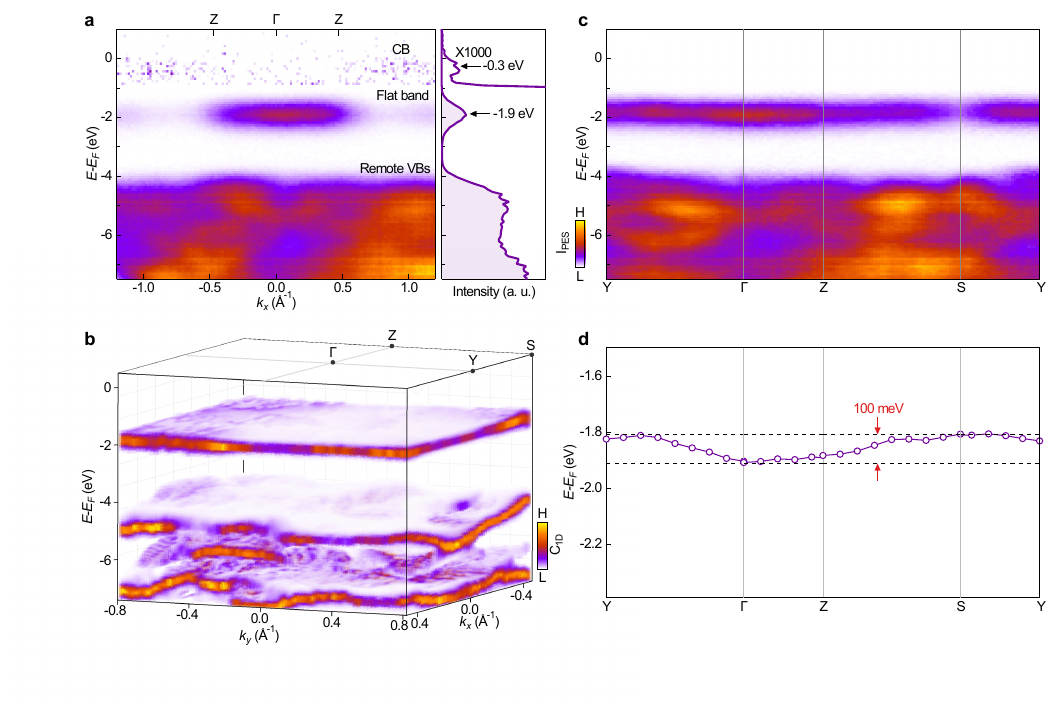}
	\caption{{\bf $\bm{|}$ Observation of an isolated flat band throughout the entire Brillouin zone.}  \textbf{a}, Dispersion image along the Z-$\Gamma$-Z direction and corresponding integrated energy distribution curve (EDC). The color scale is enhanced above -0.8 eV and the EDC is multiplied by 1000 for better visibility of the weak signal of the CB. The peak width of the flat band is larger than the energy resolution possibly owing to the intrinsic broadening or roughness of the cleaved surface. \textbf{b}, Curvature-filtered dispersion map in 2D momentum space. The color scale indicates photoelectron intensity processed by the 1D curvature filter along the energy direction. \textbf{c}, Full dispersion images along all high-symmetry momentum directions in the entire Brillouin zone. The color scale indicates photoelectron intensity in \textbf{a} and \textbf{c}. \textbf{d}, Extracted dispersion of the flat band across the Brillouin zone from the data in \textbf{c}.} 	
\label{F2}
\end{figure*}

To experimentally reveal the electronic band structure of NbOCl$_2$, a time-of-flight photoelectron momentum microscope with an extreme ultraviolet light source\cite{suga2021momentum,wallauer2021tracing} is utilized. The system can directly map the band structure over the entire Brillouin zone. It is designed for subcycle experiments with high time resolution, and the energy resolution is close to the Fourier transform limit of 180 meV. The measured dispersion image along the $\Gamma$-Z direction, together with an integrated energy distribution curve (EDC), is shown in \ref{F2}a. A flat band is clearly observed at an energy of -1.9~eV below the Fermi level, with other more remote VBs occurring below -4~eV. By enhancing the color scale, very weak signals of the CB are identified with a peak position at -0.3~eV lying above the flat band (see Supplementary Figures 1 and 2 for more data). Based on the observed CB, a single-particle band gap of 1.6 $\pm$ 0.1 eV is obtained. The much weaker intensity compared to the VB may be due to the photoemission matrix element effect\cite{damascelli2003angle} or a non-uniform distribution of carrier density. We also note that the CB is only observed in the second Brillouin zone (Supplementary Figure 3), possibly due to the momentum dependence of the matrix element of photoemission\cite{damascelli2003angle}. Compared to the similar compound NbOI$_2$\cite{abdelwahab2022giant,huang2025coupling,handa20242d}, NbOCl$_2$ exhibits a stronger Peierls distortion, which leads to much larger energy separation between the flat band and the remote VBs\cite{jia2019niobium} and makes NbOCl$_2$ an ideal system to explore flat-band physics. Therefore, a well-isolated flat band is observed with large energy separations of 1.6~eV and 2.1~eV from the CB and the remote VBs, respectively. 

The full two-dimensional (2D) band structure is mapped out in \ref{F2}b, which clearly shows that the flat band extends over the entire Brillouin zone. To explore the flatness of the band in the 2D momentum space, \ref{F2}c shows the dispersion image along representative high-symmetry directions. Clearly, the band remains almost dispersionless throughout the entire Brillouin zone. Even though the energy resolution is limited, slight variations in the energy position of the maximum of the photoemitted electrons at different momenta can still be resolved by fitting the EDCs and are confirmed in different samples (see Supplementary Figures 4 and 5 for more details). The extracted dispersion of the flat band (\ref{F2}d) reveals a bandwidth of only $\sim$100 meV, with slightly stronger (weaker) momentum dependence along $\Gamma$-Y/Z-S ($\Gamma$-Z/Y-S). The top of the flat band is located along the S-Y direction. These observations are consistent with first-principles calculations\cite{mohebpour2024origin}. Due to the weak interlayer coupling, which has been demonstrated both theoretically\cite{jia2019niobium,mortazavi2022highly,ye2023manipulation} and experimentally\cite{guo2023ultrathin}, the flat band is naturally dispersionless for out-of-plane momenta. We note that such a small bandwidth is comparable with the correlated CDW material 1T-TaS$_2$\cite{zwick1998spectral}, small-twist-angle bilayer graphene\cite{inbar2023quantum,li2024evolution}, and rhombohedral graphite\cite{zhang2024correlated}. Furthermore, it is smaller than the Hubbard U of Nb $d$ orbitals ($\sim$500 meV)\cite{moore2024high}, implying possible strong electronic correlation effects when an appropriate doping level is achieved.

\begin{figure*}[htbp]
	\centering
	\includegraphics[]{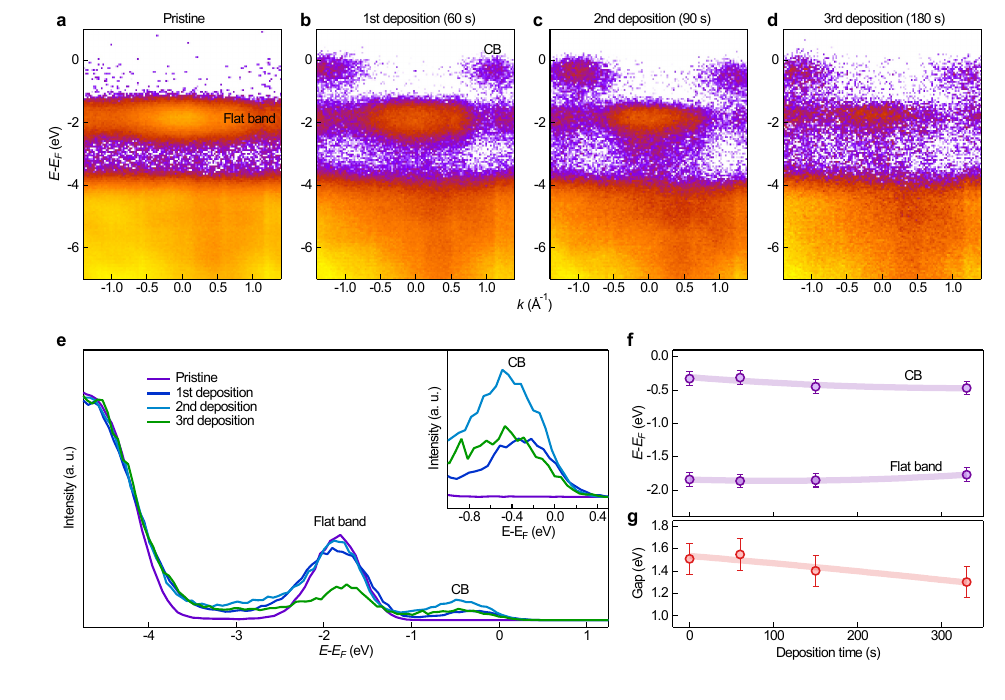}
	\caption{{\bf $\bm{|}$ Revealing the quasiparticle band gap by caesium deposition.} \textbf{a-d}, Dispersion images along the Z-$\Gamma$-Z direction at different caesium deposition stages. A logarithmic color scale is used to clearly show the CB. \textbf{e}, Integrated EDCs across momentum space from the data in \textbf{a-d}. The inset shows a zoom-in around the Fermi level, highlighting the CB. \textbf{f}, Extracted energy positions of the CB and the flat band as a function of deposition time. The thick curves are guides to the eye. \textbf{g}, Band gap extracted from the energy difference between the CB and the flat band. The thick curve is a guide to the eye.}
\label{F3}
\end{figure*}

\noindent\textbf{Band structure tuning upon caesium deposition}\\
The weak signal of the CB and the quasiparticle band gap of NbOCl$_2$ are further confirmed by $in$-$situ$ surface electron doping. By directly depositing caesium atoms on the sample surface, electrons from the adsorbed caesium atoms diffuse into the sample due to their low work function, leading to effective electron doping accompanied by an out-of-plane electric field. \ref{F3}a-c shows the dispersion images of NbOCl$_2$ upon deposition, with the signal of the CB rising for increasing deposition durations. Eventually, extensive deposition leads to suppressed CB and flat band signatures  (\ref{F3}d) due to the enhanced electron scattering on the surface. The anisotropic dispersion of the CB (Supplementary Figure 6) is consistent with theoretical calculations\cite{jia2019niobium}, which excludes any defect states as the origin of the photoelectron signal. To reveal the quasiparticle band gap, the energy positions of the CB and the flat band are extracted from the integrated EDCs in \ref{F3}e and plotted as a function of deposition time in \ref{F3}f.  The quasiparticle band gap is obtained from the energy difference between the CB and the flat band (\ref{F3}g). Surprisingly, the band gap is strongly reduced from 1.6~eV to 1.3~eV upon deposition, indicating a high band gap tunability with the carrier density and the out-of-plane electric field.

The observed band gap tunability might arise from the direct coupling between the electronic orbitals and the out-of-plane electric field\cite{kim2015observation}. Since the CB is mainly formed by the Nb 4$d_{xy}$ orbitals\cite{ye2023manipulation}, which have out-of-plane components, it can couple to the out-of-plane electric field. It may also come from negative electronic compressibility due to high-density electrons confined in the 2D surface\cite{riley2015negative}. Dual gate-controlled transport measurements can be used to disentangle the effect of the electric field and the carrier density in the future. Considering that NbOCl$_2$ can be easily exfoliated down to a few layers\cite{guo2023ultrathin}, the observed band gap tunability provides promising opportunities for controlling flat-band physics by external gating.

The observed occupation of the CB before doping is another unexpected finding, which is important for both exciton and correlation physics. Theoretical calculations  have predicted the CB to be fully unoccupied\cite{jia2019niobium,guo2023ultrathin}. The actual experimental CB occupation possibly originates from defects in the NbOCl$_2$ crystal\cite{tagani2024impact}. Recently, transport measurements have indicated the intrinsic n-type nature of as-grown NbOCl$_2$ (Ref.\cite{liu2023ferroelectricity}). 
The free carriers in the CB might screen the electron-hole interaction and prevent the formation of excitons. This indicates that carrier density tuning is important for the exploration of exciton physics in NbOCl$_2$. Moreover, to initiate strong electronic correlations in NbOCl$_2$, precise control of the carrier density is also required to shift the Fermi level towards the flat band. Doping methods such as ionic liquid gating\cite{saito2016highly,zhu2023long} can be applied. Another possible direction is the optimization of the crystal growth method to reduce defects or introduce chemical doping.

\begin{figure*}[htbp]
	\centering
	\includegraphics[]{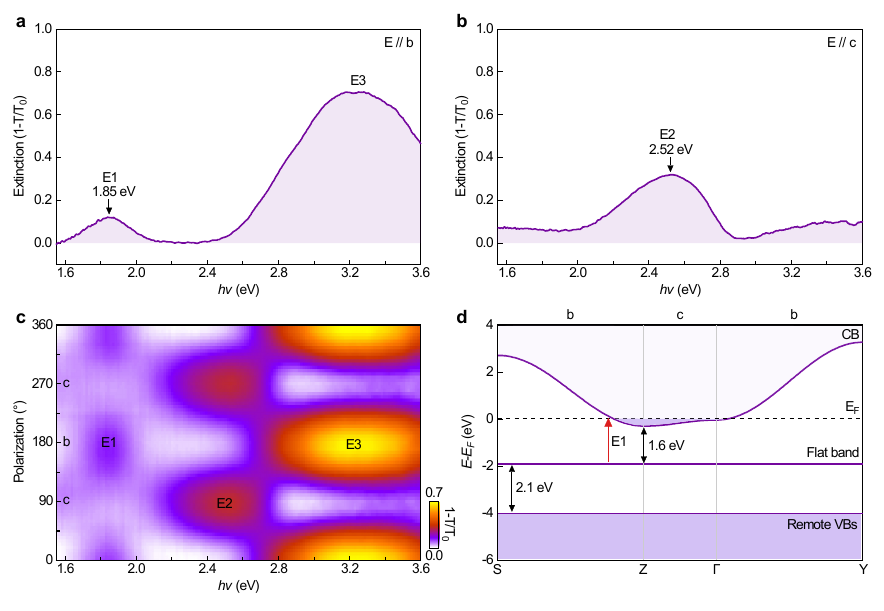}
	\caption{{\bf $\bm{|}$ Anisotropic optical gap.} \textbf{a, b}, Optical extinction spectrum for light polarized along the b-axis (\textbf{a}) and the c-axis (\textbf{b}). \textbf{c}, Optical extinction map as a function of light polarization and photon energy. The corresponding crystal orientations are labeled. \textbf{d}, Schematic band structure with experimentally determined bottom of CB, flat band, and band edge of the remote VBs. The observed optical gap is indicated by the red arrow.} 
\label{F4}
\end{figure*}

\noindent\textbf{Determination of the optical band gap}\\
The experimentally observed band structure further allows us to reveal the origin of the optical band gap. For light polarized along the b-axis, the optical transmission spectrum (\ref{F4}a) indicates a prominent absorption peak at E1 = 1.85~eV as well as a broad absorption region E3 around 3.2~eV. In contrast, light polarized along the c-axis experiences an absorption maximum at E2 = 2.52~eV (\ref{F4}b). This strong anisotropy can be seen more clearly by gradually changing the polarization in \ref{F4}c. Moreover, the absorption is independent of the light intensity (Supplementary Figure 7), suggesting a dominating role of one-photon absorption in the observed transmission spectra. Therefore, the optical gap is specified as 1.85~eV. This value is consistent with photoluminescence measurements where a broad photoluminescence peak centered at 1.9~eV is observed\cite{huang2024ferroelectric}. Due to the quasiparticle band gap of 1.6~eV and finite doping of the CB (0.3 eV), the expected optical gap would be located around 1.9~eV, as indicated by the red arrow in \ref{F4}d. This is consistent with our experimental observation.

\noindent\textbf{Conclusions}\\
In summary, by performing photoelectron momentum microscopy measurements with an extreme ultraviolet light source, the band structure of the van der Waals crystal NbOCl$_2$ is revealed. A flat band across the entire Brillouin zone is directly observed with a small width of only 100 meV. This flat band is well isolated from both the nearest CB and VB with large energy separations of 1.6 eV and 2.1 eV, respectively. Moreover, the quasiparticle band gap shows a strong tunability upon caesium deposition on the surface. By combining the band structure and optical transmission spectra, the corresponding optical transition is identified for a gap energy of 1.85 eV.

The observed flat band makes the van der Waals crystal NbOCl$_2$ an ideal system to explore the role of dispersionless quasiparticles in lightwave electronics\cite{reimann2018subcycle} and light-induced band engineering\cite{rodriguez2021low}. For example, intraband transport is expected to be strongly suppressed in the flat band owing to the large effective mass, which can lead to exotic interband-dominated\cite{li2025interband} and quantum geometry-dominated\cite{liu2025geometric}  high-harmonic generation. In addition, the ultrafast dynamics of photo-carriers in the flat bands would be distinct from common dispersive bands due to the squeezed energy space, which might trigger new developments in lightwave electronics. Since the momentum microscopy system used here is also optimized for a high temporal resolution, the demonstration of the flat band immediately opens new windows to investigate lightwave-driven electrons in flat bands and even lightwave-tailored flat bands. Meanwhile, further high-energy-resolution measurements of the band structure could be employed to explore the fine structure of the flat band, such as spin splitting\cite{mohebpour2024origin}. Therefore, our work also provides new opportunities for exploring flat-band physics.

\noindent\textbf{Methods}\\
\noindent\textbf{Photoelectron momentum microscopy measurements}\\
Photoelectron momentum microscopy measurements were performed with an extreme ultraviolet beamline. The extreme ultraviolet pulses are obtained by high harmonic generation (HHG). The output of an Yb:KGW amplifier (Light Conversion Inc., CARBIDE) operated at 50 kHz is sent into a noncollinear optical-parametric amplifier (NOPA), whose output is frequency-doubled to generate pulses centered at 400 nm. These pulses are used to drive HHG in an argon gas jet. The driving pulses are filtered out by a thin aluminum foil. The 7th harmonic at a photon energy of 21.7 eV is selected by two multilayer mirrors and is focused onto the b-c plane of the NbOCl$_2$ under an incidence angle of 70$^{\circ}$ in p-polarization.
The emitted photoelectrons are detected by a time-of-flight momentum microscope\cite{wallauer2021tracing}. The total energy resolution of the setup is better than 180 meV, and the momentum resolution is better than 0.01 \AA$^{-1}$. The samples were cleaved and measured at room temperature in an ultrahigh-vacuum chamber with a base pressure of 2$\times$10$^{-10}$ mbar. Surface electron doping was performed by $in$-$situ$ deposition of caesium atoms using an alkali metal dispenser (SAES) with a current of 5 A.

\noindent\textbf{Polarization-resolved UV-visible-NIR transmission spectroscopy measurements}\\
The broadband light was generated by a quartz tungsten-halogen lamp (Thorlabs, QTH10) and focused onto the sample with an uncoated CaF$_2$ plano-convex lens ($f$ = 50 mm). The polarization was controlled by a broadband polarizer. The transmitted light was collected by a UV-visible-NIR spectrometer (Ocean Insight, FLAME-S-XR1-ES). The commercial NbOCl$_2$ crystals (HQ Graphene Inc.) were grown by chemical vapour transport. Thin NbOCl$_2$ flake samples were exfoliated by PDMS and transferred to a diamond substrate. The transmission spectrum (T) was normalized by the transmission spectrum (T$_0$) of the substrate for each polarization. The crystal orientation is determined by polarization-dependent SHG measurements (Supplementary Figure 8). All measurements were performed at room temperature under ambient conditions.

\noindent\textbf{Curvature filtering}\\
The data in Fig.~2b is processed by a curvature filter to enhance the peak visibility\cite{ito2023build}. For this, the measured raw data is initially smoothed along the energy axis by a simple moving average (SMA) filter with a window size of 0.42~eV, which is applied three consecutive times. After this, we calculate the curvature\cite{zhang2011precise} $C(E) = \frac{f^{''}(E)}{(C_0+f^{'}(E)^2)^{3/2}}$, where we use $C_0 = 1$. As the numerical derivatives introduce a stripe-like pattern along the energy axis, we apply a 1D Median filter with a window size of 0.42~eV to remove this artefact.

\noindent\textbf{Data availability}\\
All data were processed by the Igor Pro 9.05 software. All data necessary to draw the conclusions in the paper are available within the article. All data in this study are available from Rupert Huber (rupert.huber@physik.uni-regensburg.de) and Ulrich H{\"o}fer (hoefer@physik.uni-marburg.de) upon reasonable request.

\begin{addendum}
	\item[Acknowledgements] We thank Jens G{\"u}dde, Gerd Schönhense, Anastasios Koulouklidis, Heng Zhang,  Josef Freudenstein, and Lijue Chen  for helpful discussions and Imke Gronwald, and Matthias Heinl for technical assistance. This work is supported by the Deutsche Forschungsgemeinschaft (DFG, German Research Foundation) through  SFB 1083, project-ID 223848855, SFB 1277, project-ID 314695032, research grant HO 2295/9, GRK 2905, project-ID 502572516,  DFG major instrumentation INST 89/520, project ID 445487514, and
research grant HU 1598/8, as well as by the European Research Council (ERC) Synergy Grant by project-ID 101071259. C.B. and G.I. acknowledge financial support from the Alexander von Humboldt Foundation.
	
	\item[Author Contributions] C.B., U.H., and R.H. conceived the research project. C.B., V.E., M.M., J.H., Lasse M., S.I., S.Z., G.I., and R.W. developed the momentum microscope system, performed the momentum microscopy measurements, and analyzed the data. C.B., Leon M., and L.W. performed the optical measurements and analyzed the data. C.B. and M.L. prepared the sample. C.B. and R.H. wrote the manuscript, and all authors contributed to the discussions and commented on the manuscript.	
	 
	\item[Competing Interests] The authors declare that they have no competing financial interests.

\end{addendum}

\end{document}